\documentclass[fleqn,10pt]{wlscirep}
\usepackage[utf8]{inputenc}
\usepackage[T1]{fontenc}
\title{DNA Probe Computing System for Solving NP-Complete Problems}

\author[1,+,*]{Jin Xu}

\author[2,+]{XiaoLong Shi}

\author[3,+]{Xin Chen}

\author[4]{Fang Wang}

\author[2]{Sirui Li}

\author[2]{Pali Ye}

\author[2]{Boliang Zhang}

\author[2]{Di Deng}

\author[2,*]{Zheng Kou
}

\author[2,4,*]{Xiaoli Qiang
}

\affil[1]{Key Laboratory of High Confidence Software Technologies, School of Computer Science, Peking University, No.5 Yiheyuan Road, Haidian District, Beijing, 100871, China}
\affil[2]{Institute of Computing Science and Technology, Guangzhou University, 230 Wai Huan Xi Road, Guangzhou Higher Education Mega Center, Guangzhou, 510006, China}
\affil[3]{Key Laboratory for Physical Electronics and Devices of the Ministry of Education and Shaanxi Key Laboratory of Photonics Technology for Information, School of Electronic Science and Engineering, Xi’an Jiaotong University, No.28, Xianning West Road, Xi’an 710049, China}
\affil[4]{School of Computer Science and Cyber Engineering, GuangZhou University, 230 Wai Huan Xi Road, Guangzhou Higher Education Mega Center, Guangzhou, 510006, China}

\affil[*]{corresponding.author@jxu@pku.edu.cn, kouzheng@gzhu.edu.cn, qiangxl@gzhu.edu.cn}

\affil[+]{these authors contributed equally to this work}

\begin{abstract}
Efficiently solving NP-complete problems—such as protein structure prediction, cryptographic decryption, and vulnerability detection—remains a central challenge in computer science. Traditional electronic computers, constrained by the Turing machine’s one-dimensional data processing and sequential operations, struggle to address these issues effectively. To overcome this bottleneck, computational models must adopt multidimensional data structures and parallel information processing mechanisms. Building on our team’s proposed probe machine model (a non-Turing computational framework), this study develops a \textbf{blocking probe technique} that leverages DNA computing’s inherent parallelism to identify all valid solutions for NP-complete problems in a single probe operation. Using the 27-vertex 3-coloring problem as a case study, we successfully retrieved all solutions through DNA molecular probe experiments. This breakthrough demonstrates the first implementation of a \textbf{fully parallel computing system} at the molecular level, offering a novel paradigm for tackling computational complexity. Our results indicate that the probe machine, with its parallel architecture and molecular implementation, transcends the limitations of classical models and holds promise for solving intricate real-world problems.
\end{abstract}
\begin{document}

\flushbottom
\maketitle
%
%
\thispagestyle{empty}


\section*{Introduction}

Computational tools have driven human progress since antiquity, from primitive tally systems to modern electronic computers\cite{haigh2016eniac}. While innovations like mechanical calculators and transistor-based architectures revolutionized speed and precision, the linear logic of Turing machines remains ill-suited for high-dimensional combinatorial optimization such as NP-complete problems\cite{arora2009computational}.

Over the past five decades, scientists have pursued unconventional computing paradigms—including quantum, optical, and biological computing—to surmount these limitations. In 1959, Feynman’s conceptualization of molecular computing \cite{lambert19591959}inaugurated a new era of untraditional computation. A pivotal milestone emerged in 1994 when Adleman exploited the intrinsic parallelism of DNA molecules to solve combinatorial optimization problems\cite{1adleman1994molecular}. Subsequent research has systematically advanced DNA encoding optimization\cite{2tanaka2005design}, computational model design\cite{3roweis1996sticker}\cite{4winfree1998design}\cite{5benenson2001programmable}, and implementation technologies\cite{6liu1999progress}\cite{7sakamoto2000molecular}\cite{8cukras1999chess}, culminating in innovations such as DNA-based finite-state automata\cite{9benenson2001programmable}, programmable DNA computers\cite{11braich2002solution}, and DNA origami-enabled visual computing systems\cite{10woods2019diverse}.

Nevertheless, the engineering practice of molecular computing necessitates resolving three core challenges: First, what questions can molecular computers best answer? Second, which of the many tools used by molecular biologists can best promote molecular computation? And third, how can the architecture of a molecular computer best be designed? DNA computing has demonstrated unique advantages in solving NP-complete problems, including the 7-vertex Hamiltonian path problem\cite{1adleman1994molecular}, 20-variable 3-SAT problem\cite{11braich2002solution}, maximum clique problem\cite{12ouyang1997dna}, and 61-vertex graph coloring\cite{13xu2018dna}. DNA’s parallelism makes it an ideal medium for constructing exponentially large solution spaces. However, traditional DNA computing remains bottlenecked by solution-space screening inefficiencies—multi-step serial operations (e.g., gel electrophoresis, enzymatic purification) introduce latency, inherently adhering to the Turing machine’s sequential logic. The critical challenge lies in rapidly isolating valid solutions from the vast solution space, as the reliance on serial experimental workflows perpetuates the constraints of Turing-style computation.

In the research of biomolecular computing, our research group has proposed a novel underlying fully parallel Probe Machine (PM) model\cite{14xu2016probe}. The Probe Machine is composed of nine core components: Data Library, Probe Library, Data Controller, Probe Controller, Probe Operation, Computing Platform, Detector, True Solution Storage, and Residue Collector. It denoted as

\begin{equation}
PM=\left(X, Y, \sigma_{1}, \sigma_{2}, \tau, \lambda, \eta, Q, C\right).\label{eq1}
\end{equation}

Different from the Turing machine, this architecture inherently integrates storage and computation, enabling arbitrary related information processing to be completed simultaneously with data reading. It represents a novel paradigm distinct from the von Neumann architecture. To implement the theoretical model of the Pobe Machine, our research group designed an innovative blocking probe technique for the 3-coloring problem of graphs—a typical NP-complete problem. Through a single probe operation, all solutions to the problem within a certain scale were obtained, and finally, high-throughput nanopore sequencing analysis was used to validate the experimental results. This is the first non-Turing computing model proposed by humanity, breaking through the one-dimensional data processing mode of traditional Turing Machines and providing a new approach for the research and development of novel computing models and chips. Such innovation not only offers fresh perspectives for tackling complex computational challenges but also opens broad prospects for advancing future computational science.

\begin{figure}[ht]
\centering
\includegraphics[width=\linewidth]{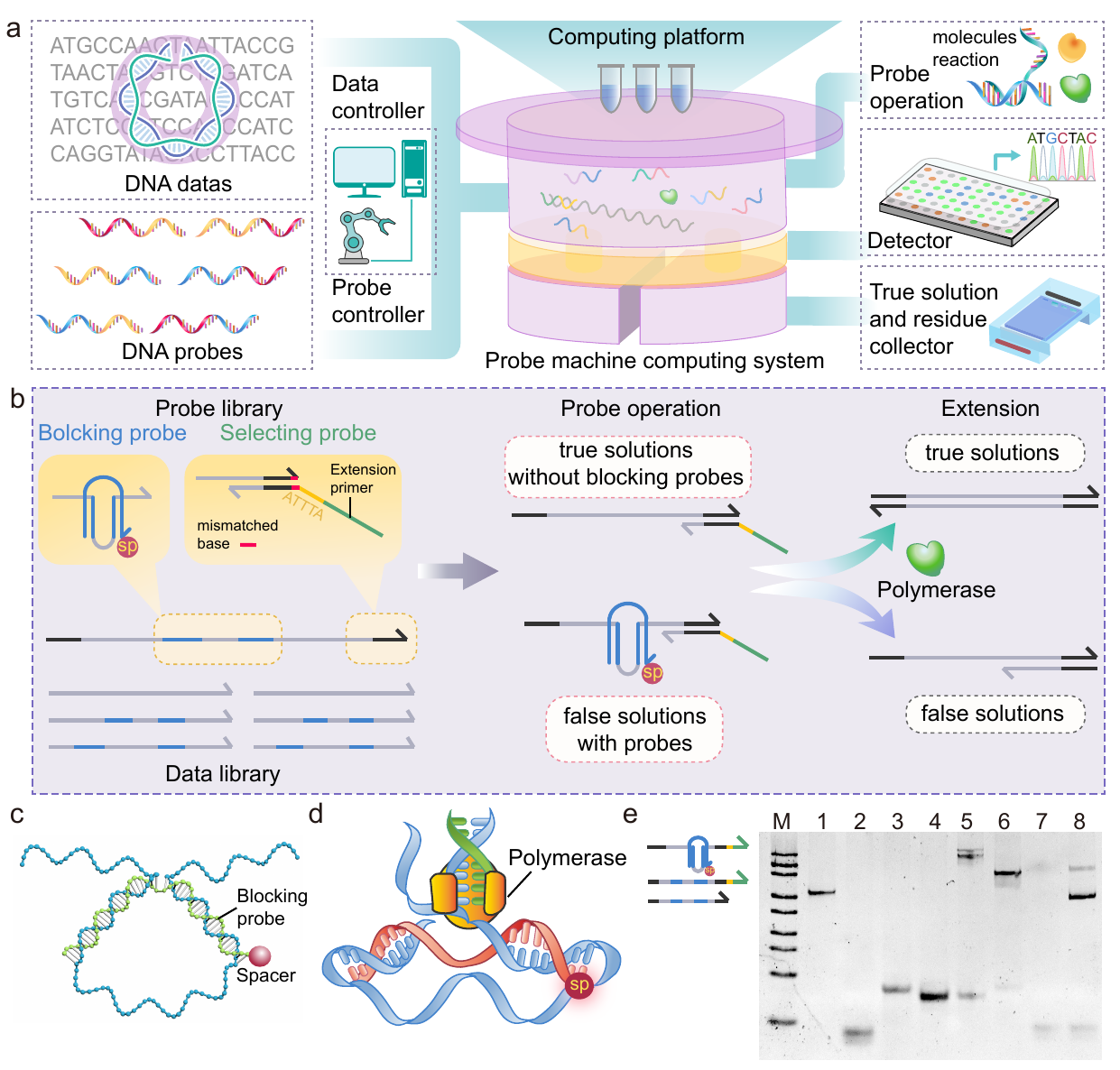}
\caption{\textbf{The probe computing system.} (a) Mathematical model of the probe computing system—nine-tuple Probe Machine (PM) model (b) Schematic diagram of the working principle of the probe computing system (c) Schematic diagram of the blocking probe structure (d) Schematic diagram of the extension process in the probe computing system
(e) System verification experiment diagram. Lane 1: Single-stranded solution space;Lane 2: Forward and reverse primers; Lane 3: Selection probe; Lane 4: Blocking probe; Lane 5: Extension with the participation of both selection and blocking probes; Lane 6: Extension with the participation of only the selection probe; Lane 7: Amplification after extension with the participation of both selection and blocking probes; Lane 8: Amplification after extension with the participation of only the selection probe.}
\label{fig1}
\end{figure}

\section*{Probe Computing System}

DNA demonstrates inherent parallelism and programmability, making it an exceptional material for solving NP-complete problems. However, current implementations of DNA parallelism primarily excel in constructing exponential solution spaces containing both valid and invalid solutions. The elimination of invalid solutions still relies on sequential experimental operations constrained by the Turing machine paradigm. The probe computing system, grounded in an underlying fully parallel probe machine model, enables single-step elimination of invalid solutions through blocking probes (Fig. 1a). Within this system, blocking probes and selection probes operate synergistically. With DNA polymerase participation, the process unfolds in two stages: extension and amplification, achieving precise segregation of invalid solutions. During extension, both valid and invalid solutions undergo guided elongation via selection probes. Invalid solutions are arrested upon recognition by blocking probes, enabling subsequent separation based on molecular weight disparities after extension (Fig. 1b).

The selection probe consists of three parts. The first part is complementary to the partial sequence before the last base at the 3' end of the possible solution. The second part contains 5 free bases, which ensures that the last base at the 3' end of the solution remains free and prevents the 3' end of the solution from extending towards the third part. The third part is the valid - solution identification sequence, which is used for the exponential amplification of valid solutions during the recovery process. As a guiding primer to assist the blocking probe in selecting valid solutions, the selection probe ensures that unblocked valid solutions can complete full - length extension. Then, using the valid - solution identification sequence primer and the forward primer pair, exponential amplification of the fully - extended valid solutions is carried out. In contrast, invalid solution sequences that cannot complete full - length extension under the action of the blocking probe cannot undergo exponential amplification in the next stage.

The blocking probe consists of two recognition sites. The 3' end of the blocking probe is modified with a spacer, which is used to stop PCR. After the blocking probe recognizes the target strand, it binds to the target strand in the opposite direction, forming a stable paperclip - like structure (Figure 1c). Once the structure is formed, the three - line defense of the blocking system starts to function. The first line of defense: The paperclip - like structure formed by the blocking probe and the target strand shows a significant difference in molecular weight from non - target strands. The second line of defense: The 5' end of the blocking probe forms a reverse structure with the complementary end of the target strand. The extension process occurs at 37 °C, at which the activity of DNA polymerase is relatively low. The selection probe starts to extend from the 3' end of the target strand. When it reaches this reverse site, due to the sudden change in the helical pattern of the blocking probe at this location, a steric hindrance is formed, which interferes with the work of some DNA polymerases, preventing them from extending to a sufficient length (Figure 1d).  The third line of defense: After the DNA polymerase breaks through the first two lines of defense, it will stop extending when it encounters the spacer modified at the 3' end of the blocking probe.

With the cooperation of the blocking system and the primer system, polyacrylamide gel electrophoresis (PAGE) experimental analysis shows (Figure 1e) that the blocking probe forms a stable target structure with the target strand, and new bands with significantly different molecular weights appear. After the extension, compared with the control group, the bands shift significantly upwards after the addition of the blocking probe. By referring to the bands in the control group, the target bands of the two groups were recovered. Bidirectional PCR was performed on the recovered substrates using the upstream and downstream primers respectively. After the probe blocking operation, no target product was obtained, while the control group normally produced the target product. This demonstrates that the probe computing system has the ability to accurately recognize and block the extension of the target strand.

\begin{figure}[ht]
\centering
\includegraphics[width=\linewidth]{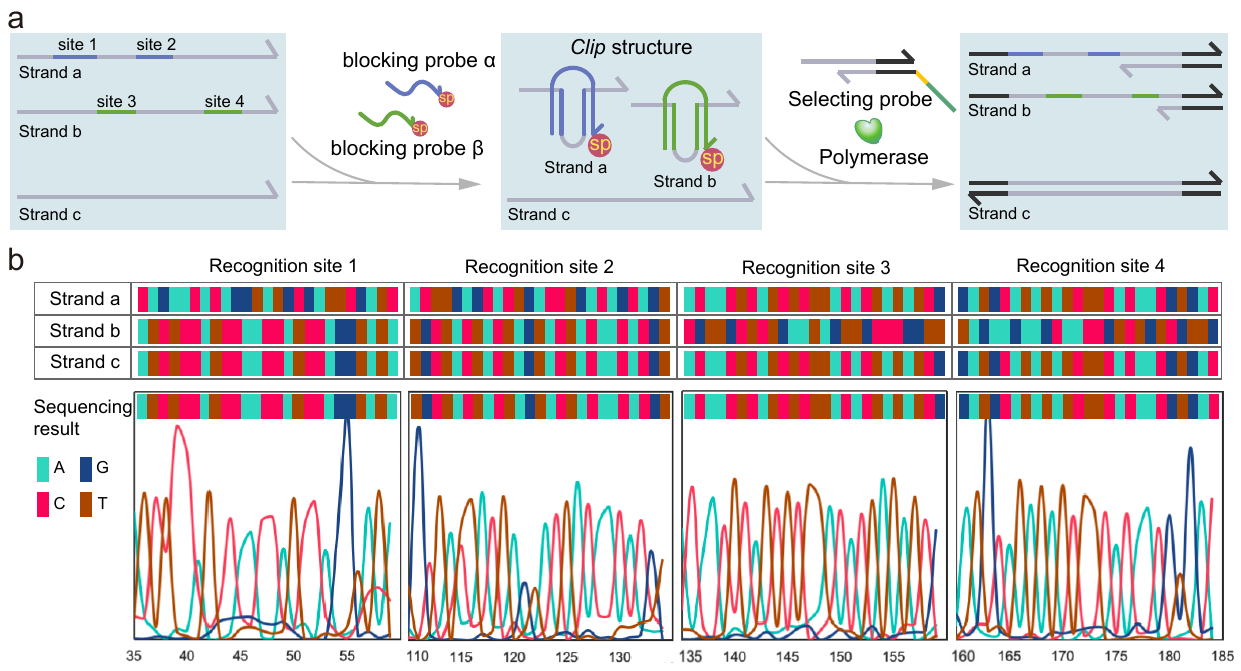}
\caption{\textbf{Verification of parallel processing in the probe computing system.} (a) Schematic diagram of the experiment on the principle of parallel processing (b) Results of Sanger sequencing. The upper table shows the sequence information of each recognition site on the three strands, and the lower part shows the sequencing results of these recognition sites.}
\label{fig2}
\end{figure}

\section*{Parallel Processing of the Probe Computing System}

To verify the parallel processing ability of the probe computing system, a solution space containing three single - stranded solutions was constructed for the verification experiment. Each solution sequence was 350 base pairs (bp) long and contained 14 sites. Strands a and b each corresponded to a blocking probe. Recognition sites 1 and 2 corresponded to blocking probe $\alpha$, and recognition sites 3 and 4 corresponded to blocking probe $\beta$. Strand c did not contain any recognition sites for blocking probes (Figure 2a). The three single - stranded solutions were mixed at equal concentrations. Blocking probes at a concentration three times that of the single - stranded solutions and selection probes at a concentration equal to that of the single - stranded solutions were added. During the annealing process, as the temperature decreased, the blocking probes in the solution freely and specifically bound to the single - stranded solutions to form stable paperclip - like structures. Under the action of DNA polymerase, the selection probes guided the single - stranded valid solutions to form double strand DNA (dsDNA) of 380 base pairs (bp). For invalid solutions, due to the steric hindrance formed by the paperclip - like structures, which was difficult for the polymerase to overcome, only shorter dsDNA structures could be formed. PAGE experiments showed that blocking probes $\alpha$ and $\beta$ bound to strands a and b respectively, forming distinct migration bands. There were also differences in the sizes of the bands of the paperclip - like structures formed by different blocking probes and single strand DNA (ssDNA). It was inferred that this was because there were obvious differences in the sizes and shapes of the formed paperclip - like structures. Using the band of strand c alone as a control, the target band was excised from the gel and PCR was performed using upstream and downstream primers. The PCR products were purified and subjected to Sanger sequencing. The results showed that only valid solutions were present in the products. By comparing the sequences of strand a, b, and c, at the positions where the sequences differed, the sequencing results showed a single peak and the sequence was completely consistent with that of strand c (Figure 2b).

On one hand, the experiment demonstrates that the blocking probes of the probe computing system can accurately recognize target strands, confirming their specificity; on the other hand, the system can simultaneously identify multiple target strands to achieve blocking effects, indicating certain parallelism. Specificity serves as the foundation for solving problems, while parallelism provides the possibility of solving large-scale problems in one step. Here, the three-coloring problem for a 27-vertex graph was solved based on the probe computing system.

\begin{figure}[ht]
\centering
\includegraphics[width=\linewidth]{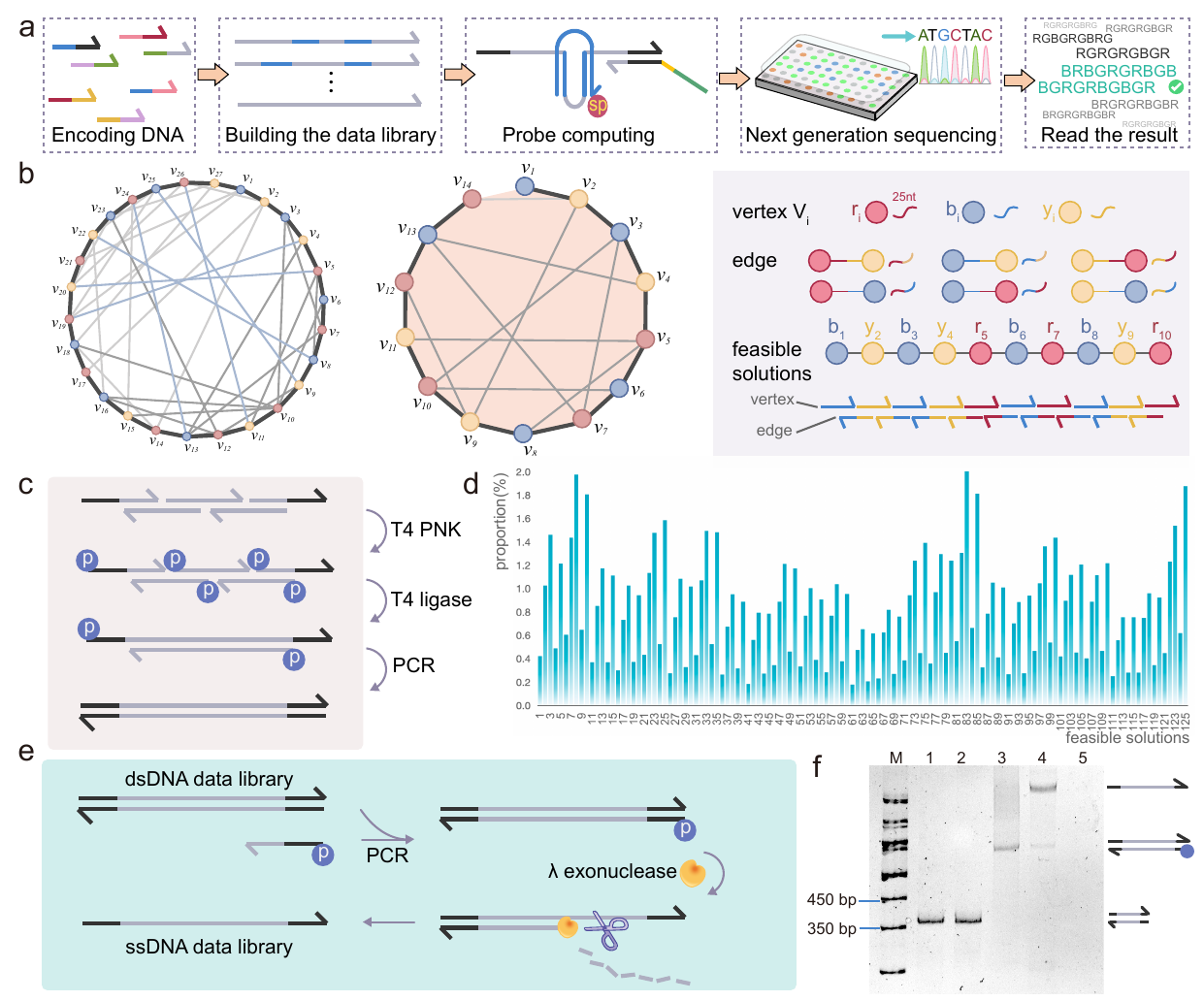}
\caption{\textbf{Construction of Solution space.} (a) Flowchart for solving large-scale problems based on the probe computing system (b) On the left is a graph $G$ with 27 vertices; in the middle is a subgraph $G_1$ of the 27-vertex graph; on the right is a schematic diagram of the encoding of DNA solution sequences (c) Schematic diagram of the construction of the double-stranded solution space (d) Sequencing alignment results of the double-stranded solution space (e) Schematic diagram of the construction of the single-stranded solution space (f) Schematic diagram of PAGE analysis for solution space construction. Lane 1: Double-stranded solution space of the half-library $G_1$; Lane 2: Double-stranded solution space of the half-library $G_2$; Lane 3: Double-stranded solution space of the full-library $G$; Lane 4: Single-stranded solution space of the full-library $G$; Lane 5: Single-stranded solution space of the full-library $G$ after treatment with S1 enzyme.}
\label{fig3}
\end{figure}

\section*{Construction of Solution Space}
The 3 - coloring problem of a graph is a classic NP - complete problem. Given an undirected graph $G=(V, E)$, where V is the set of vertices and E is the set of edges. The 3 - coloring problem of a graph aims to determine whether it is possible to color each vertex in the graph with 3 colors, such that any two adjacent vertices (i. e., vertices connected by an edge) have different colors. In this paper, the 3 - coloring problem of a graph with 27 vertices is solved based on the blocking probe computing system. The main steps of the solution include: construction of the solution space, probe operation, and reading out the correct solutions through third - generation sequencing (Figure 3a).

The first step in using DNA computing to solve mathematical problems is to map the mathematical problem onto DNA and construct a solution space. Each vertex $V_i$ in the undirected graph G corresponds to three different coloring possibilities, which are represented by three 25-nt-long DNA oligonucleotides $\{r_i, y_i, b_i\}$. An edge E connecting two vertices $V_i$ and $V_j$ is constructed using the complementary sequence of the last 13 nt of the $V_i$ sequence and the complementary sequence of the first 13 nt of the $V_j$ sequence.

When constructing the solution space, a Hamiltonian path of the graph is selected as the solution space. Theoretically, there are nine different edge sequences between two vertices. However, since adjacent vertices cannot have the same color, edges with the same color between adjacent vertices on the Hamiltonian path can be excluded, i. e., $\overline{b_i b_j}, \overline{r_i r_j}, \overline{y_i y_j}$, leaving a total of six different connecting edges (Figure 3b). The free combination of all vertex sequences and edge sequences forms the double-stranded solution space for this problem. However, vertices that are adjacent in the graph but not on the Hamiltonian path may still have the same color, thus forming a feasible solution set containing non-solutions. To form a uniform and stable double-stranded solution space and reduce interactions between edge and vertex sequences, a strategy of constructing the solution space in stages was adopted. The solution space was divided into two half-libraries by length: $G_1=\{V_1... V_{14}\}$ and $G_2=\{V_{14}... V_{27}\}$. Constructing each half-library involved four steps: T4 kinase activation, annealing, T4 ligase ligation, and amplification (Figure 3c). The two half-libraries $G_1$ and $G_2$ were then bridged to form the complete full-library $G$ (Figure 3f, Lanes 1 and 2). Notably, unlike PCR with a single template, interactions between templates during half-library and full-library construction can generate nonspecific products. To produce purer target products, PCR was performed at lower template concentrations, with real-time monitoring of fluorescence changes to stop amplification at the late exponential growth phase and enter the extension stage. Both high template concentrations and excessive PCR cycles can lead to nonspecific products. PAGE confirmed the production of pure target products (Figure 3f, Lane 3), and nanopore sequencing analysis showed that the constructed full-library had good uniformity (Figure 3d), providing a foundation for subsequent stable computations.

In the blocking probe computing system, blocking probes need to bind to ssDNA. Therefore, high - yield and high - purity ssDNA are necessary prerequisites for probe operations. The experiment attempted two different technical approaches, LATE - PCR technology and $\lambda$ exonuclease treatment, to generate ssDNA. In the initial stage of the experiment, an attempt was made to generate ssDNA using only one dsDNA as a template. In the LATE - PCR scheme, different template concentrations and ratios of primers to substrates failed to yield target products with high yield and high purity. $\lambda$ exonuclease can selectively digest 5'-phosphorylated double - stranded DNA in the 5'→3' direction, and has relatively low enzymatic activity on single - stranded DNA and 5'-unphosphorylated DNA. In the $\lambda$ exonuclease treatment scheme, the template DNA needs to be phosphorylated first. Bidirectional PCR is performed using an upstream primer and a phosphorylated downstream primer. Then, ssDNA can be obtained after treating the phosphorylated dsDNA with $\lambda$ exonuclease (Figure 2e). The single - stranded feasible solution set containing invalid solutions was constructed using the full - library G as a template with $\lambda$ exonuclease. As the amount of $\lambda$ exonuclease increases, the digestion speed also increases; as the digestion time increases, the degree of digestion also increases. However, a longer digestion time is not always better. Prolonged digestion can lead to a decrease in ssDNA yield due to over-digestion.

\begin{figure}[ht]
\centering
\includegraphics[width=\linewidth]{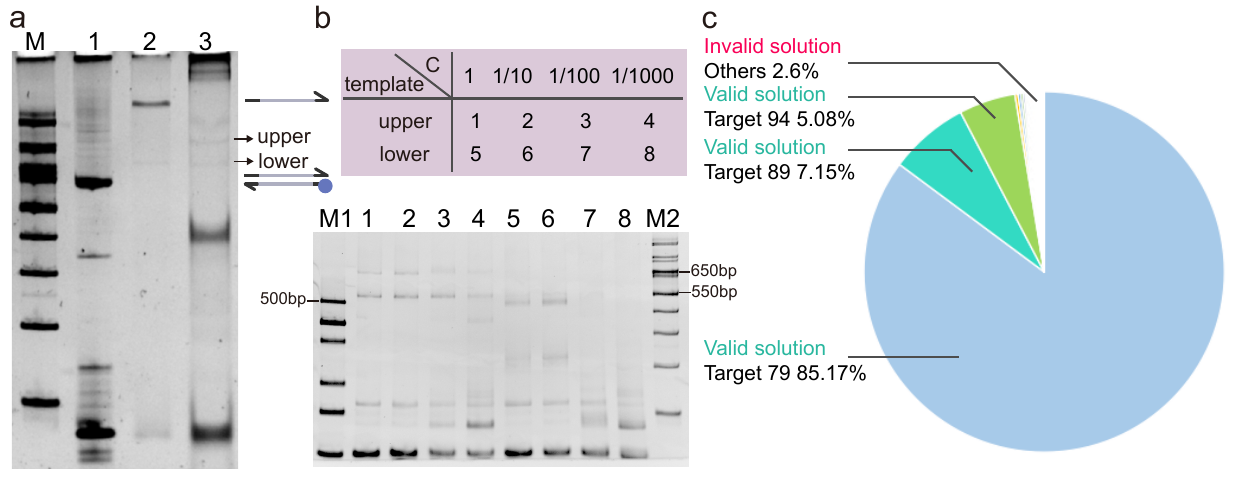}
\caption{\textbf{Results of solving large-scale problems.} (a) PAGE analysis chart of the extension results. Lane 1: Double-stranded solution space G; Lane 2: Single-stranded solution space; Lane 3: After blocked extension. (b) Amplification results after blocked extension. Lanes 1 - 4: The templates are the products recovered by excising the “upper” band in Lane 3 of Figure 4a, with concentrations being the undiluted recovered solution, 1/10, 1/100, and 1/1000 of the original respectively; Lanes 1 - 5: The templates are the products recovered by excising the “upper” band in Lane 3 of Figure 4a, with concentrations being the undiluted recovered solution, 1/10, 1/100, and 1/1000 of the original respectively. (c) Sequence alignment results after third-generation sequencing.}
\label{fig4}
\end{figure}

\section*{Blocking Probe Computing for Large-Scale Problems}

In the solution space constructed according to the Hamiltonian path, although the situation of adjacent vertices on the path having the same color is excluded, there are still cases where non - adjacent vertices on the path but adjacent in the graph have the same color. Therefore, the constructed single - stranded solution space contains non - solutions, and these non - solutions need to be removed to obtain the correct solutions for the graph. According to the adjacency relationship of vertices in the graph, the complementary sequences of the sequences of two adjacent vertices with the same color are the blocking probes. For example,$\overline{r_5 r_8}$. Solving the graph with 27 vertices requires 15 blocking probes to remove non - solutions. In the experiment, annealing, extension, and symmetric PCR were carried out in sequence according to the probe system operation method introduced above. However, the results of PAGE showed that the band after extension was slightly higher than the double - stranded band. It was inferred that most of the sequences in the solution space were extended, but the blocking probes failed to block the dsDNA. The experiment verified this inference using third - generation sequencing. The sequencing results showed that the sequence distribution was relatively uniform. Compared with the sequencing results of the library strands, the types of sequences were similar, and although the distribution ratios were slightly different, they were basically similar.

To investigate the reasons for the failure of the probe computing system, an attempt was made to generate multiple solution sequences by combining vertex sequences according to the size of the solution space. These sequences were analyzed using NUPACK under physiological conditions (37°C). The results revealed that a single solution strand can form numerous hairpin structures on its own, and when multiple solution strands coexist, they can also match each other to form secondary structures. Therefore, as the problem scale increases, the solution space becomes more complex, and the similarity between sequences becomes higher. Additionally, the annealing process promotes the interwinding of solution strands to form complex secondary structures. These intertwined secondary structures block the binding of blocking probes to solution strands, preventing the formation of blocking structures and causing the probe computing system to fail. To address this, the annealing process was replaced by increasing the concentration of blocking probes. Eliminating the temperature gradient of annealing reduces the possibility of forming large secondary structures, while high probe concentrations promote binding between blocking probes and solution strands.

When solving large-scale problems, the blocking efficiency of the probe computing system was improved from another perspective by reducing the activity of DNA polymerase. During the extension process, high-fidelity DNA polymerase was used to ensure extension accuracy, but higher activity would increase the blocking pressure on the probe computing system. Before extension, the high-fidelity enzyme was not subjected to heat activation, and extension was completed using only the enzyme's activity before heat activation. Experiments showed that the yield of target bands in the experimental group without heat activation treatment was significantly higher than that in the heat-activated group.

During repeated experiments, the target band at the expected position was often indistinct. However, in the control group without any blocking probe recognition sites, after recovering the band at the corresponding position and performing symmetric PCR with upstream and downstream primers, the target product of the corresponding length was obtained. We infer that this is because the symmetric PCR efficiency of a single template during the library construction stage is much higher than that of complex multi-type templates. This issue was resolved by concentrating the large-volume PCR reaction solution using a PCR recovery kit to increase its concentration, making the target band clear and easy to excise and recover from the gel. After recovery, upstream and downstream primers were used to amplify the product.

After adopting the strategy of "low temperature, low enzyme activity, and high initial concentration", PAGE analysis results after extension showed (Figure 4a) that in addition to sequences bound to probes appearing at higher positions, two bands existed below the single-strand band position and above the double-strand position. These bands were excised and recovered from the gel separately, and amplified using different substrate concentrations. The amplification results (Figure 4b) showed that the upper band produced an amplified band of the expected length, while the lower band failed to amplify correctly. Therefore, the upper band was considered the result of the operation.

\section*{Verification of Solutions}

By adopting the strategy of "low temperature, low enzyme activity, and high initial concentration", the solution to the graph with 27 vertices can be obtained after the operation of the probe computing system. The experiment used third - generation sequencing technology to detect the amplified products. This is a high - throughput detection method with single - molecule resolution. The DNA strands in the amplified products repeatedly pass through the protein pore arrays on the membrane surface in a short time, and the sequence information is output based on the differences in the current signals generated by different bases blocking the pores. In this way, the DNA sequence components of the solution library can be detected.

To verify the accuracy of the calculation results of the probe computing system, a silicon-based computer was used to construct the solution space of the graph with 27 vertices. The BLAST algorithm was employed to perform sequence matching between the solution space and the calculation results. After being processed by graph theory algorithms, the solution space was greatly simplified. The solution space constructed by the silicon-based computer had a total of 125 possible solutions, among which 3 were correct solutions.The sequencing results (Figure 4c) contained 32,528 valid data entries. After BLAST alignment, it was shown that although there were sequences matching 68 possible solutions, the number of matches for 65 of these possible solutions was less than 100, and for 41 of them, the number was even less than 10. The number of the 3 possible solutions that met the requirements of the correct solutions was 31,681, accounting for 97.40\% of all the data. However, the distribution of the three correct solutions was not uniform, with proportions of 85.17\%, 7.15\%, and 5.08\% respectively. The uneven proportion of the sequences of the three correct solutions might be due to differences in PCR efficiency caused by sequence variations.

\section*{Discussion}

In this paper, a probe computing system based on the probe machine model was constructed to solve graph coloring problems. The solution to the problem can be obtained through solution space construction, probe blocking-extension operation, amplification, and sequencing. The programmability of DNA is utilized to convert graph information into DNA sequence information, and a non-enumerative model is used to specifically and parallelly construct the solution space. The probe computing system sets three lines of defense, using only one DNA oligonucleotide to accurately identify non-solution strands. All non-solution blocking probes are added at once according to non-solution generation conditions to eliminate all non-solutions. The third-generation nanopore sequencing technology is employed to rapidly obtain problem solutions in a short time. Taking a graph with 27 vertices as an example, this solving method fully leverages the programmability, parallelism, and low-energy consumption characteristics of DNA computing to successfully solve search problems with a scale of $3^{25}$.

The parallelism of the probe computing system greatly simplifies the serial experimental operations of traditional biological computing methods constrained by the "Turing machine" model, enabling solutions to be obtained within polynomial time with an algorithmic complexity of O(n). The probe computing system has scalability: on one hand, the system itself has the potential to expand the scale of parallel solving; on the other hand, by combining methods such as graph theory subgraph partitioning, it can solve subgraphs in parallel to expand the scale of solvable problems.

The graph coloring problem is an intuitive combinatorial optimization problem with broad applicability. The constraint that "adjacent vertices cannot share the same color" can simulate various complex real-world scenarios. For example, in traffic scheduling, vertices can represent tasks with time conflicts in different regions; in resource allocation, adjacent regions can signify allocation restrictions for competitive resources. Precisely for this reason, the graph coloring problem has become a universal model for many practical issues, ranging from protein structure prediction to wireless spectrum allocation, all of which can be solved by transforming them into graph coloring problems.

The 3-coloring problem is a typical NP-complete problem. Stephen A. Cook revealed a fundamental law in the field of computational complexity: all NP-complete problems are equivalent to each other in polynomial time. He also constructed a unified reduction framework – through polynomial-time transformation, any NP problem can be mapped to an instance of another NP-complete problem. This discovery greatly simplifies the paradigm of computational complexity research: researchers don't need to tackle each NP problem one by one. Instead, they can indirectly solve the entire NP problem domain by studying representative problems with strong modeling capabilities, such as the graph coloring problem.

In summary, this study solved the "graph coloring" problem through blocking probe technology, verifying the feasibility of the probe machine model in solving NP-complete problems. Experiments show that the combination of DNA molecular parallelism and specificity can break through traditional computational bottlenecks, laying a theoretical foundation for the engineering of biomolecular computers. Future work will focus on automated probe design, development of hybrid computing architectures, and their applications in NP-complete problems, particularly protein structure prediction.

\section*{Methods}

\subsection*{DNA Oligonucleotides} All DNA oligonucleotides were purchased from Sangon Biotech Co. Ltd. (Shanghai, China). Unmodified DNA strands were purified via poly- acrylamide gel electrophoresis and dissolved in ddH2O (Sangon Biotech Co., Ltd, China). Modified DNA strands were purified using high-performance liquid chromatography and dissolved in sterilized ddH2O (Sangon Biotech Co., Ltd, China).

\subsection*{PCR Amplification} The concentrations of forward and reverse primers are 1 $\mu$M, and the substrate is generally a 1/1000 dilution of the recovered sample. 2X SYBR Green Abstart PCR Mix (Sangon Biotech Co., Ltd, China) was used for PCR. The program was performed on a QuantStudio 3 (Thermo Fisher Scientific Inc., USA) as follows: heat activation at 95°C for 2 minutes, followed by cycles of 95°C for 10 seconds and 62°C for 2 minutes until the end of the exponential phase, and finally an extension at 72°C for 5 minutes.

\subsection*{Polyacrylamide Gel Electrophoresis} The DNA solutions were analyzed in an 5\% native polyacrylamide gel in 1X TAE buffer after running for 50 min(375 bp)or 105 min(675 bp) at a constant power of 100 V using electrophoresis apparatus: EPT-VPG04 (ZHONGKEYONGYI CO., LTD., China). Gels were stained with GelRed, which was purchased from Sangon Biotech Co., Ltd. (Shanghai, China), and scanned with a scanner (ProteinSimple, USA).

\subsection*{Third-generation nanopore sequencing} The rapid library preparation kit (Qitan Technology, China) was used to construct the sequencing library, with a sample input of 220 fmol. After library construction, the QNome - 3841 nanopore gene sequencer (Qitan Technology, China) was used to read the sequence information continuously for 20 hours. Use the BLAST algorithm to align sequences.

\section*{Funding}

National Natural Science Foundation of China [62172114, 62473104, 62072129, 62332006]; National Key R\&D Program of China [2019YFA0706402].

\section*{Conflict of interest statement}

None declared.






 




\bibliography{sample}

\end{document}